\documentclass{article}
\usepackage{ijcai17}

\usepackage{times}

\usepackage{algorithm}
\usepackage{graphicx}
\usepackage{amsmath}
\usepackage{algpseudocode}
\usepackage{nccbbb}
\usepackage{tabu}
\usepackage[small]{caption}
\usepackage{flushend,cuted}
\usepackage{balance}

\title{Sequential Prediction of Social Media Popularity with \\Deep Temporal Context Networks}

\author{Bo Wu$^{1,2}$, Wen-Huang Cheng$^{3}$, Yongdong Zhang$^{1,2}$, Qiushi Huang$^{4}$, Jintao Li$^{1}$, Tao Mei$^{5}$\\
$^{1}$Institute of Computing Technology, Chinese Academy of Sciences, China\\
$^{2}$University of Chinese Academy of Sciences, China\\
$^{3}$Academia Sinica, Taiwan\\
$^{4}$University of Surrey, UK \\
$^{5}$Microsoft Research, China\\
\{wubo, zhyd, jtli\}@ict.ac.cn; whcheng@citi.sinica.edu.tw; hqsiswiliam@gmail.com; tmei@microsoft.com
}

\begin{document}
\setlength\titlebox{2.5in}
\maketitle

\begin{abstract}
Prediction of popularity has profound impact for social media, since it offers opportunities to reveal individual preference and public attention from evolutionary social systems. Previous research, although achieves promising results, neglects one distinctive characteristic of social data, i.e., sequentiality. For example, the popularity of online content is generated over time with sequential post streams of social media. To investigate the sequential prediction of popularity, we propose a novel prediction framework called Deep Temporal Context Networks (DTCN) by incorporating both temporal context and temporal attention into account. Our DTCN contains three main components, from embedding, learning to predicting. With a joint embedding network, we obtain a unified deep representation of multi-modal user-post data in a common embedding space. Then, based on the embedded data sequence over time, temporal context learning attempts to recurrently learn two adaptive temporal contexts for sequential popularity. Finally, a novel temporal attention is designed to predict new popularity (the popularity of a new user-post pair) with temporal coherence across multiple time-scales. Experiments on our released image dataset with about 600K Flickr photos demonstrate that DTCN outperforms state-of-the-art deep prediction algorithms, with an average of 21.51\% relative performance improvement in the popularity prediction (Spearman Ranking Correlation).
\end{abstract}
 

\begin{figure*}
  \centering
  \includegraphics[width=7.0in]{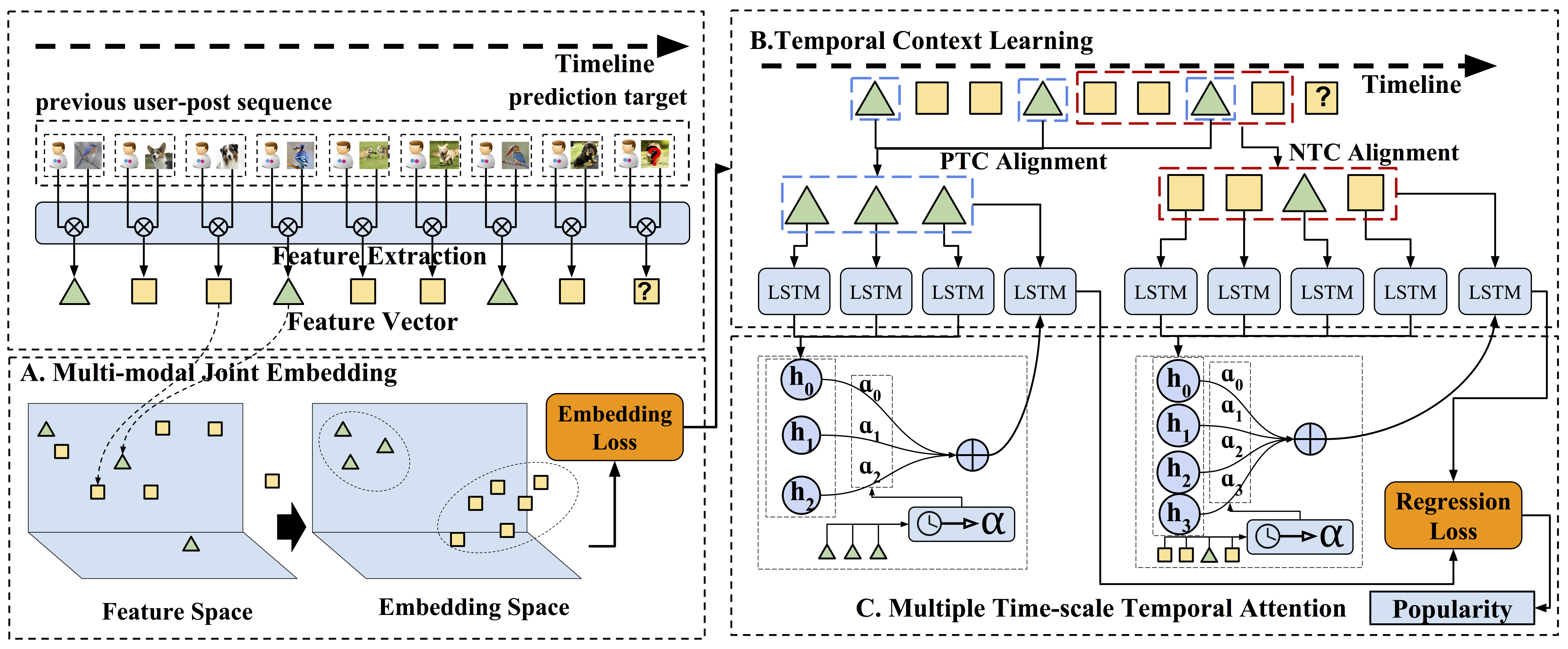}
  \caption{Overview of Deep Temporal Context Network (DTCN). (A) Multi-modal Joint Embedding converts user feature and visual feature into an embedding space, i.e., mapping the two kinds of multi-modal information into a same latent space; (B) Temporal Context Learning constructs two temporal contexts, and learns contextual information based on LSTM; (C) Multiple Time-scale Temporal Attention assists the final prediction process based on temporal attention mechanism. (Best view in color) }\label{framework}

\end{figure*}

\section{Introduction}
Social media is now globally ubiquitous and prevalent. Consequently, understanding and predicting popularity in social media (e.g., Twitter, Facebook, Youtube and Flickr) has attracted great attention~\cite{Wu2016Unfolding,Wu2016Temporalization,Li2015,He2014,Khosla2014,Pinto2013}, since it offers opportunities to reveal individual preference and public attention from evolutionary social systems. Accurate popularity prediction can help improve user experience, service effectiveness, and benefit a broad range of applications, such as content recommendation~\cite{Khosla2014}, online advertising~\cite{Li2015} and information retrieval~\cite{Roy2013,Gan2016}.

Previous research, although achieves promising results, neglects one distinctive characteristic of social data, i.e., temporal sequentiality. For example, the popularity generated by photo sharing in Instagram on a weekday is observed to be at the peak during hours at 2 a.m., 8-9 a.m. and 5 p.m.~\cite{Beese2016instagram}. Existing predictive algorithms on social media popularity are not considering the temporal order of data, making them have limited success to sequential data scenarios (e.g. news feed, tweet timeline, photo stream, etc.). Particularly, most of the existing works on popularity prediction are based on the content of a post or the person who published the post~\cite{Shamma2011,Gelli2015,Cappallo2015}. Recently, although some researchers have analyzed temporal characteristics of social popularity (e.g., temporal fluctuations), the temporal interrelationship of popularity data is not explicitly exploited~\cite{Shen2014,Zhao2015,Wu2016Temporalization,Wu2016Unfolding,Martin2016}.

In this research, we take one step further to investigate the problem as a temporal prediction task with sequentiality. Unlike previous work on using time information as latent factors or variables~\cite{Shen2014,He2014,Zhao2015}, our purpose is to predict popularity from both sequential and temporal views based on time series data. Specifically, we incorporate a time-centered perspective called `temporal context' into popularity prediction, which was inspired by social psychology for human behavioral processes~\cite{McGrath1992}. Utilizing temporal context of popularity as a novel prior knowledge, we attempt to predict new popularity by exploiting consecutive temporal coherence of popularity at multiple time-scales.


Based on the above idea, we propose two types of sequential temporal contexts for learning two different types of temporal coherence of popularity: Neighboring Temporal Context (NTC) and Periodic Temporal Context (PTC). On one hand, we model NTC to learn the continuous trend or patterns in short-term time series. For instance, before and after Thanksgiving Day, the popularity of pumpkin picture rises and decays in only a few days. The NTC helps us reveal the trend and variances. Such temporal coherence has also been successfully applied to the citation estimation of scientific articles~\cite{Shen2014}. On the other hand, we model PTC to learn discontinuous temporal coherence in long-term time series. Temporal coherence is often influenced by periodic events or human activities. For example, according to the survey in 2016~\cite{Ellering2016}, the best time periods to make a post on Facebook are Saturday and Sunday around 12 a.m. to 1 p.m., with periodic peaking time from 9 a.m. to 3 p.m. And weekly peak time for Pinterest is on Saturday from 8 p.m. to 11 p.m. These findings motivate us to consider both of the temporal and sequential characteristics for predicting popularity more precisely. 

In this paper, therefore, we propose a novel deep prediction framework called Deep Temporal Context Networks (DTCN) by exploring both temporal contexts and temporal attention at different time-scales jointly (such as days of a week, hours of a day). Figure ~\ref{framework} shows the overview of our DTCN framework, which is a sequential prediction architecture containing three main components: from embedding, learning to predicting, along with the model trained as an entire network by optimization learning. With a joint embedding network, we map multi-modal data onto the same embedding space to obtain a unified deep representation for user-post sharing activities. Then, based on the embedded data sequence, we design temporal context learning to recurrently learn the dynamic popularity from adaptive temporal contexts (NTC and PTC) for popularity over time. Finally, we provide a multiple time-scale attention for computing multi-scale temporal coherence in predicting the popularity of a new user-post pair. Unlike previous work on temporal modeling for popularity prediction~\cite{Wu2016Unfolding,Wu2016Temporalization,Li2015}, our study attempts to provide a novel view on temporal context modeling, which considers both of the temporal and sequential coherence during prediction.

The main contributions of this study are: (i) To our best knowledge, we are the first to consider both temporal and sequential characteristics into sequential prediction of social media popularity over time; (ii) we address the problem by incorporating a temporal context perspective, and proposing two types of novel temporal contexts including NTC and PTC; (iii) we propose a novel deep prediction model DTCN architecture jointly integrating embedding, temporal context learning and predicting with temporal attention to optimize the entire network, which outperforms state-of-the-art predictive algorithms in sequential popularity prediction.

\section{Related Work}
Recently, time-aware popularity prediction receives much attention. Both academia and industry have paid more effort on this research topic. Existing works about prediction with temporal information can be concluded into two main paradigms.
The first paradigm focuses on predicting the popularity growth of a published post by analyzing its temporal trend and pattern at early stage. Szabo and Huberman proposed to predict popularity based on growth pattern characteristics of online popularity at early stage~\cite{Szabo2010predicting}. Yang and Leskovec found the temporal patterns reveal how the content’s popularity grows and fades during the post propagation~\cite{Yang2011}. Roy \emph{et al.} proposed to grasp sudden bursts of a post popularity with cross-domain knowledge~\cite{Roy2013}. Kong \emph{et al.} explored the problem of detecting which hashtag would be bursting. Most of these works provided dynamic comprehension on popularity accumulation of online content, but their prediction needs to rely on early stage popularity patterns of a published post. Our method takes one step further on popularity prediction before that the corresponding post was published. 

The other paradigm is predicting popularity based on temporal features or dynamic signals. Zhao \emph{et al.} applied human reaction time as temporal variables in self-exciting point processes~\cite{Zhao2015}. Shen \emph{et al.} proposed a reinforced Poisson process to model the dynamic popularity based on the arrival time of attention~\cite{Shen2014}. He \emph{et al.} designed a time-aware bipartite graph for popularity estimation~\cite{He2014}. Wu \emph{et al.} proposed to predict popularity by context factorization and tensor decomposition algorithms to unfold popularity dynamics~\cite{Wu2016Unfolding,Wu2016Temporalization}. These models all neglect the sequentiality of popularity during prediction.

\textbf{Summary.} We focus on the investigation of sequential temporal coherence of popularity over time before the corresponding sharing behavior happened. Being differed from previous works on the predicting popularity with temporal information e.g.,~\cite{Shen2014,He2014,Zhao2015,Wu2016Temporalization,Wu2016Unfolding}, we explore the temporal context of popularity from sequential data and consider both of sequential and temporal characteristics of popularity during prediction. It is worth mentioning that our algorithm is generic and applicable to the prediction of other sequential data scenarios in social media, e.g. news feed, tweet timeline, photo stream, and so forth.

\section{Social Media Popularity Prediction}
Taking photo popularity prediction on Flickr as an exemplary case, we define our problem and introduce the definitions of several core concepts, including user-post sequence, temporal context, and multiple time-scales. 

\textbf{Problem Definition}: Given a new photo $v$ of a user $u$, the problem of predicting its popularity $s$ is to estimate how many attentions would be obtained after the post was published on social media (e.g. views, likes or clicks etc.).

On Flickr, when browsing a personal photo stream or image search results, users can view details of a photo content with its metadata through clicking photo thumbnails. In our prediction, since ``viewing count'' is a significant indicator of how popular a photo is, we use it to describe the photo popularity as follows:

\textbf{Popularity Normalization.} To suppress the large variations among different photos (e.g. view count of different photos vary from zero to millions), we implement a log function to normalize the value of popularity, based on the previous work~\cite{Wu2016Temporalization,Khosla2014}. In brief, the log-normalization function for popularity can be defined as:
 \begin{equation}
 s=log_{2}{\frac{r}{d}}+1,
 \end{equation}
where $s$ is the normalized value, $r$ is the view count of a photo, and $d$ is the number of days since the photo was posted.

Several core concepts in the paper are also defined as follows:

\textbf{Definition 1: User-Post Sequence.} A user-photo pair  $p \Leftrightarrow \left\langle {u, v} \right\rangle$ is derived from the photo sharing behavior (e.g., post publishing, photo sharing and video uploading), which made by user $u$ on photo $v$. Suppose we have $n$ user-photo pairs and the sharing time of each pair. Then the user-post sequence can be denoted by $S = \{ \left\langle {u_1, v_1} \right\rangle,\left\langle {u_2, v_2} \right\rangle ,...,\left\langle {u_n, v_n} \right\rangle \}$ with its sharing time order $t_{1} \leq t_{2} \leq ... \leq t_{n}$.

\textbf{Definition 2: Temporal Context.} Given the prediction target $p$ and its previous user-post sequence $P_u$ before time $t$, temporal context (as a prior knowledge) is a time series, which was built on $P_u$ and expressed by a triple sequence: $C_p = \{ \left\langle {u_1, v_1, t_{p_1}} \right\rangle,\left\langle {u_2, v_2, t_{p_2}} \right\rangle ,...,\left\langle {u_k, v_k, t_{p_k}} \right\rangle \}, t_{p_1}\leq t_{p_2} \leq ... \leq t_{p_k} \leq t$ where $k$ is the item count in the temporal context.

\textbf{Definition 3: Multiple Time-scales.} For time order sequence, time-scale determines a time unit of the sequential data. In our paper, without loss of generality, there are four levels of time-scales, $T_{unit} = \{t_{1M}, t_{1P}, t_{1D}, t_{1W}\}$, which means that minute of a hour, period of a day, day of a week and week of a month. With regard to the period of a day, we segment one day into six periods~\cite{Wu2016Temporalization}, i.e. ``morning (8:00am-12:00am)'', ``lunch time (12:00am-14:00pm)'', ``afternoon (14:00am-17:00pm)'', ``dinner time (17:00am-20:00pm)'', ``evening (20:00am-24:00pm)'' and ``night (0:00am-8:00am)''.

\section{Deep Temporal Context Network}
How can the temporal context at multiple time-scales be utilized for popularity prediction? We address the problem as a sequential prediction task, where the input is a user-photo sequence (with time order) while the output is the popularity of a ``future'' photo (a photo before its publication on social media). We propose a deep prediction framework called DTCN containing three components: (1) Multi-modal Joint Embedding (2) Temporal Context Learning and (3) Multiple Time-scale Temporal Attention. With all the components, the model is trained as an entire network by optimization algorithm RMSprop~\cite{Tieleman2012}.

\subsection{Multi-modal Joint Embedding}
In the first stage of DTCN, Multi-modal Joint Embedding (MJE) is to generate a unified deep representation of multi-modal sequence data. While an individual popularity is highly correlated with user-photo interactions~\cite{Cappallo2015,Khosla2014,Qian2014}, our embedding model is not only designed to correctly understand what appears in photos but also incorporate the knowledge of who posts it. 

Suppose that a variety of photo sharing behaviors over time can be naturally viewed as a user-photo sequence. The bimodal sequence data are the input data of our embedding network. In order to parametrize the visual content and user influence from the data sequence, we design a two-stream Feed-forward Neural Network (FNN), which has both of the user analysis pipeline and the photo analysis pipeline respectively. On one hand, the photo analysis pipeline starts with a pre-trained ResNet~\cite{He2015} in 152 layers for generating high-level visual representations of 2048 dimensions. On the other hand, we adopted a group of features to measure user influence in the user analysis pipeline, such as the average value of views, photo count, the number of contacts, mean number of members in a user's groups, and having a Pro Flickr account or not. In our embedding network, each of the pipelines contains four layers with two hidden layers and shares the same dimension number of hidden neural nodes of 256 and 32. In order to perform non-linear mapping of features from the original space to a new latent space, the $tanh$ activation function is applied in each layer of the embedding networks. Meanwhile, we apply a random dropout mechanism with parameter 0.5 in each hidden layer of the two pipelines, since dropout strategies have been adopted to prevent over fitting in neural network training phrase~\cite{Srivastava2014dropout}. In the end of the architecture, user and photo information $x_{u}$ and $x_{v}$ generated from the last layers of the two-stream FNN are embedded together by minimizing on embedding loss:
 \begin{equation}
 L(x_{u},x_{v})=\sum_{x_{u}\in U,x_{v}\in V} || softmax( x_{u}\odot x_{v}-x_{u}\odot x_{v} )||_{2}^2\\,
 \end{equation}
where $U$ and $V$ are the collections of all users and photos respectively, and $||\cdot||_{2}$ means L2 norm regularization for loss function. The final output of the embedding network is a 64-dimensional vector. By joint training with all the other components of the DTCN framework, we obtain an embedding representation of the user-photo sequence.

\subsection{Temporal Context Learning} 
In the learning stage of DTCN, Temporal Context Learning (TCL) is to learn sequential and temporal coherence from the temporal context for prediction. Different from traditional context learning, our temporal context is adaptive time series and correlated with the time information of a prediction target and its previous user-photo sequence.

The first step of TCL is constructing temporal contexts as prior knowledge for each ``future'' popularity. To model short or long-term fluctuations of popularity over time, we propose two variable-length temporal contexts: \emph{Neighboring Temporal Context} (NTC) and \emph{Periodic Temporal Context} (PTC). Suppose we already have the embedding data of previous user-photo sequence $P_u$ for a prediction target $p$, the data items are corresponding with the user-photo pairs in $P_u$. Specifically, the temporal context $C_p$ is a time series generated from the time information of $P_u$. Therefore, NTC consist of neighboring items of the prediction target from its previous data sequence, which is applied to describe rapidly-varying popularity fluctuations in a short-term time range; PTC is built with discontinuous 
s from the previous data sequence, which is applied to represent periodic popularity patterns in a long-term time range. Here we design two discrimination functions of NTC and PTC to determine whether a certain item $s \in P_u$ would be a possible element of the corresponding temporal context. The NTC discrimination function is defined as follows: 
 \begin{equation}
 f_{NTC}(t_p,t_s) = \delta(\frac{t_p-t_s}{\Delta t_{unit}}<l),  \forall s\in P_u, 
 \end{equation}
where $t_p$ and $t_s$ is the sharing time of $p$ and $s$, and $\Delta t_{unit}$ is the duration of a time unit. Dirac delta function $\delta(\cdot)$ is to compute a discrimination signal, and we use $l$ in discrimination function to control the signal condition lengths of time span for temporal context computation. In addition to short-term fluctuations, popularity also has periodic variances in long-term. Therefore the discrimination function for PTC is:
 \begin{equation}
 f_{PTC}(t_p,t_s)  = \delta (\bmod(\frac{t_p-t_s}{\Delta t_{unit}}=0)),  \forall s\in P_u 
 \end{equation}
 where the modulo function $\bmod (\cdot)$ is to detect whether $p$ and $s$ are in the same time blocks in different time-scales (e.g. the same period in different days or the same day in different weeks). In order to learn consecutive variance from both of NTC and PTC, we maintain the time-scale $t_{unit} \in T_{unit}$ of them to be same in context learning.

To learn consecutive coherence from temporal contexts, we provide a two-stream recurrent neural network based on LSTM~\cite{Hochreiter1997long} with Mean Squared Error (MSE) loss function as the optimization objective. Our network not only considers contextual information but also utilizes temporal information to learn the temporal context for both short-term and long-term temporal coherence. Taking the embedding data sequence as the input of LSTM, TCL is able to learn and read the surrounding temporal context for each prediction target by controlling recurrent state updates of the network.

\subsection{Multiple Time-scale Temporal Attention}
Multiple Time-scale Temporal Attention (MTTA) is proposed to consider the dynamic impact of each contextual item into prediction. Intuitively, we incorporate the attention mechanism to be \emph{infer} the temporal attention across multiple timescales between previous instances and new instances for new popularity prediction. Different from traditional attention, multiple timescale coherent among data need to be learnt from both of weights of relative hidden states and multiple scales of time-series.

Considering all data items of temporal context, the temporal context vector $c_i$ is the input of LSTM at the step $i$. It can be computed by a weighted sum of hidden state $h_i$:
 \begin{equation}
 c_i=\sum_{j=1}^{|C_{p}|}\alpha_{ij}h_{j},
 \end{equation}
where the $\alpha_{ij}$ is attention weight. It is calculated by comparing current prediction target $p$ and the data item of previous temporal context at position $j$:
 \begin{equation}
 \alpha_{ij}=\frac{\exp(e_{ij}^{-1})}{\sum_{k=1}^{|C_{p}|}\exp(e_{ik}^{-1})},\\
 e_{ij}=1-\frac{\bar{t}_i\cdot\bar{t}_j}{||\bar{t}_i||^2||\bar{t}_j||^2},
 \end{equation}
where $\bar{t}$ denotes multi-scale time vector. Note that unlike in the traditional attention mechanism, we use attention score $ e_{ij}$ to leverage the temporal consistency between $\bar{t}_{p}$ and $\bar{t}_{s}$. Its calculation relies on time vector $\bar{t}=(t_{1M}, t_{1P}, t_{1D}, t_{1W})$ with  multiple time-scale information instead of hidden state vector $h$ for temporal attention. To compute the temporal consistency, we apply cosine distance function as a simple metric, while it can be alternated by other distance functions. 

\section{Experiments}
In this section, we demonstrate the effectiveness of the proposed framework on Flickr dataset as follows: (1) we compare performances between our proposed method and current state-of-the-art algorithms for popularity prediction on different data scales respectively. (2) we provide the experiment results of using single temporal context on DTCN and demonstrate the performance results with different context types, considering temporal unit and time-scales. 

\subsection{Experimental Setup}

\textbf{Temporal Popularity Image Collection (TPIC17)\footnote{We released the data set in https://github.com/social-media-prediction/flickr-data-prediction-2017}.} TPIC17 is an image popularity dataset with multi-faceted information, such as user profile, photo metadata, and visual content. To construct the sequential prediction dataset and protect the privacy of photo sharing behaviors, we extracted time information of the adopted time-scales from the original timestamps of photo sharing. It contains 680K photos in total, and the sharing time of photos are over three years from Flickr. In order to use the temporal information from the dataset, we extracted time information from the metadata. To obtain multiple data settings with different dataset size, we individually sampled three sub datasets (100K, 200K and 400K) from the 680K dataset to evaluate algorithms.

\textbf{Moving Partition Validation.} In order to evaluate sequential prediction task, we have a 5-round moving partition strategy on the segment of training and testing data for validation. We organized the entire data in time order and divided it into 14 parts in total. The moving partition strategy is using data from a moving time window recurrently. In our experiments, we use the data in time window as input for each round. In our experiments, the length of time window are 10, which means the partition ratio of training versus testing data is 9:1.

\textbf{Evaluation Metrics.} We evaluated the prediction performance of our approach and the baselines on a correlation metric Spearman Ranking Correlation (SRC) and a precision metric Mean Absolute Error (MAE). SRC is to measure the ranking correlation between ground-truth popularity set $P$ and predicted popularity set $\hat P$, varying from 0 to 1. If there are $k$ samples, the SRC can be expressed as:
\begin{equation}
r_s = \frac{1}{k-1} \sum ^k _{i=1} \left( \frac{P_i - \bar{P}}{\sigma_P} \right) \left( \frac{{\hat{P}}_i - \bar{\hat{P}}}{\sigma_{\hat{P}}} \right), 
 \end{equation}
where $\bar{P}$ and $\sigma_P$ are mean and variance of the corresponding popularity set. Furthermore, we also use 
Mean Absolute Error (MAE) to calculate the averaged prediction error: 
\begin{equation}
MAE=\frac{1}{k}\sum_{i=1}^{n}\mid\hat{P}_{i}-P_{i}\mid.
 \end{equation}

\begin{table}[tb]
\centering
\caption{Prediction performances on TPIC17-100K, 200K, and 400K datasets (metric: Spearman Ranking Correlation).}
\label{SRC Comparison}
\begin{tabular}{cccc}
\hline
\textbf{Dataset}           & \textbf{100K} & \textbf{200K} & \textbf{400K} \\ \hline
\textbf{Metric} &       \multicolumn{3}{c}{\textbf{SRC}}            \\ \hline
CNN-AlexNet                & 0.2450               & 0.2435               & 0.1423               \\
CNN-VGG                    & 0.2281               & 0.2445               & 0.1064               \\
SVR                        & 0.2647               & 0.2367               & 0.2289               \\
SVR(T)                     & 0.2741               & 0.2984               & 0.2643               \\
MLP                        & 0.4135               & 0.5282               & 0.5559               \\
MLP(T)                     & 0.4436               & 0.5068               & 0.5084               \\
LSTM                       & 0.4629               & 0.5837               & 0.5885               \\
CLSTM                      & 0.4966               & 0.5730               & 0.6072               \\
\textbf{DTCN}              & \textbf{0.5990}      & \textbf{0.6175}      & \textbf{0.6692}      \\ \hline
\end{tabular}
\end{table}

\subsection{Compared Methods}
In order to compare with state-of-the-art models, we implemented the following approaches which can be applied into popularity prediction task as baselines.

\emph{Baseline 1 \& 2:} \textbf{Convolutional Neural Networks (CNN-AlexNet~\cite{krizhevsky2012imagenet} and CNN-VGG~\cite{ILSVRC15}).} CNNs have been proved a powerful tool in the field of image understanding for photo popularity prediction ~\cite{Cappallo2015}. Consequently, we apply CNN-AlexNet and CNN-VGG as the baselines for CNNs, and fine tuned the 8-layer AlexNet and 19-layer VGG as a regression task that is optimized with our image dataset to predict photo popularity. Different from other methods, the input of CNNs baseline is original image file only. 

\emph{Baseline 3:} \textbf{Support Vector Regression (SVR)~\cite{Khosla2014}.}
Khosla~\emph{et.al.} implement SVR algorithm in popularity prediction task, which utilized user information and visual content together as feature vectors by using linear kernel. Simultaneously, we add temporal information into SVR as another baseline method that denotes as SVR(T).

\emph{Baseline 4:} \textbf{Multiple Layer Perceptron (MLP)~\cite{zhang1998comparison}.} MLP is a typical feedforward neural network trained with back propagation, which has the general ability and does not require any assumption about the distribution of training data to solve real-world problems. Meanwhile, we also implement a variant MLP(T) by adding temporal information to the input vector of MLP.

\emph{Baseline 5:} \textbf{Long Short-Term Memory (LSTM) \cite{Hochreiter1997long}.} LSTM is a recurrent neural network (RNN) architecture, which is capable of dealing with sequential information for popularity prediction. 

\emph{Baseline 6:} \textbf{Contextual LSTM (CLSTM)~\cite{ghosh2016contextual}.} Since the contextual deep learning is a relatively new research direction, there are only few existing research works to compare with. To the best of our knowledge, the most closely related work is CLSTM, which also considered contextual information in training and predicting phrase. Therefore, we used Contextual LSTM as a representative of Contextual Recurrent Neural Network (CRNN). Different from traditional LSTM, CLSTM takes the contextual information into account during the training and predicting process, and it has been proved feasible in NLP tasks. The parameters of Recurrent Layers (e.g. LSTM) and hidden neural nodes shared the same settings over proposed prediction framework (DTCN), LSTM and CLSTM. The numbers of output dimensions are 64 with the ``hard sigmoid'' activation function.

\begin{figure}
  \centering
  \includegraphics[width=3.2in]{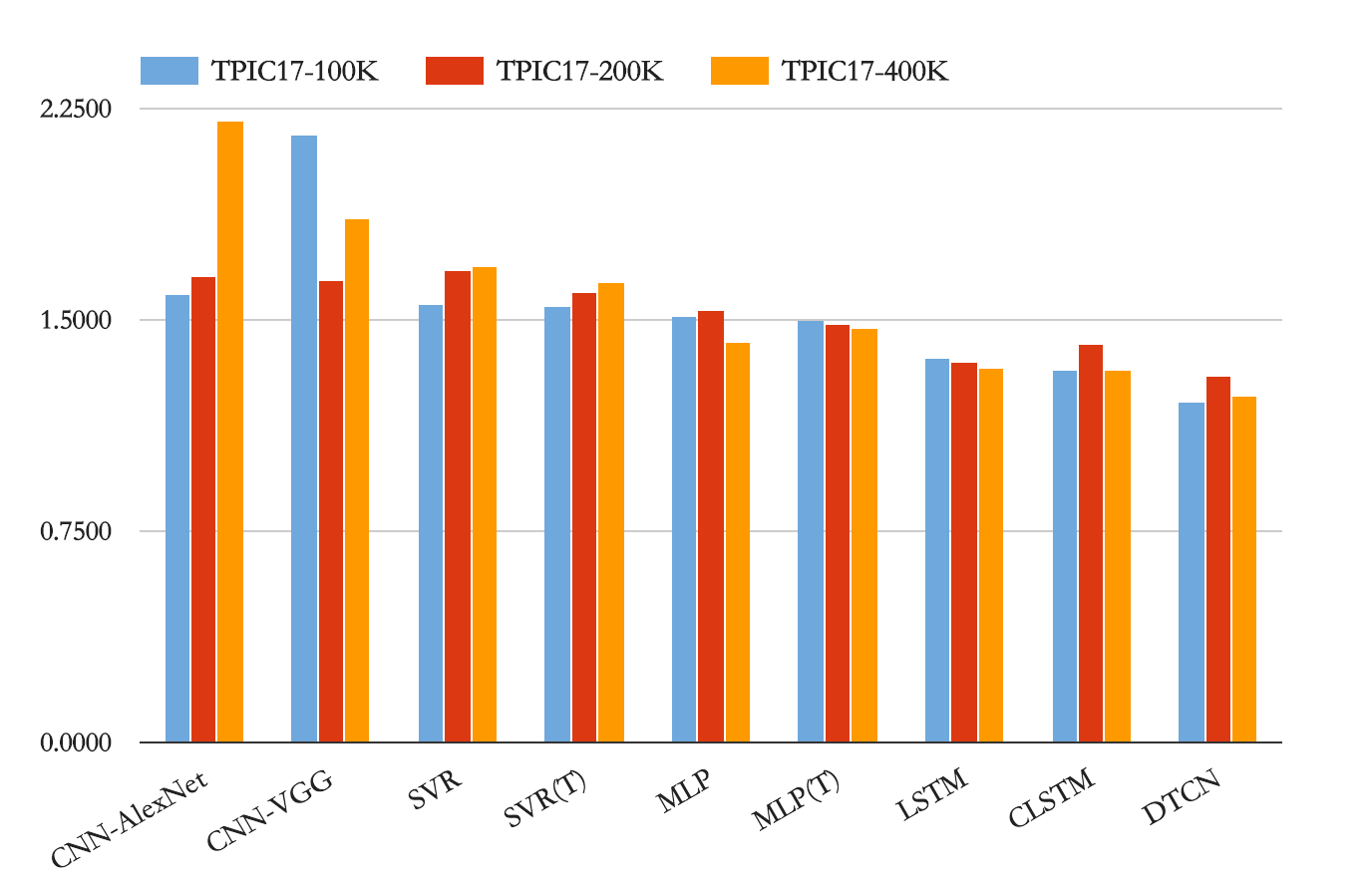}
  \caption{MAE (Mean Absolute Error) comparison of different algorithms on 100K, 200K and 400K datasets.}\label{MAE Comparison}
\end{figure}

\subsection{Prediction Performance}
The prediction performances of our proposed method and the compared models are shown in Table \ref{SRC Comparison} and Figure \ref{MAE Comparison}. Overall, our approach DTCN achieves the best prediction performance on three data size settings with highest SRC of 0.6692 and minimum MAE 1.2341. DTCN outperforms state-of-the-art deep prediction algorithms (MLP, MLP(T), LSTM and CLSTM), with an average of 21.51\% relative performance improvement on SRC. Using raw data from images directly without user metadata and temporal information, both CNN-AlexNet and CNN-VGG receive the worse results compared to other deep learning baseline algorithms. As multiple layer networks, MLPs obtains higher effectiveness than SVRs on prediction, while both of them are using user and visual features. From SVR and MLP to SVR(T) and MLP(T), there are neck-to-neck results across different sizes of datasets. 
Thus involving temporal information as non-sequential feature only has limited power for prediction, the behind reason might be ignorance of the sequential coherence of temporal information. 
With consideration of sequential coherence in the modeling of prediction, LSTM and CLSTM achieve good performances on SRC, i.e., 0.4629 and 0.6072, which are better than MLP and lower than DTCN. Overall, the results of sequential prediction (i.e. RNNs) are much superior than the results of non-sequential prediction (e.g. SVR, MLP or CNNs) on predicting social media popularity. The prediction accuracies of different methods are shown in Figure ~\ref{MAE Comparison}. From the histogram of MAE, our model also achieves the minimal prediction error 1.2341. From CNNs to DCTN, it can be seen that the error drops from about 2.1 to 1.2, and DTCN has stable improvements on TPIC17-100K, TPIC17-200K, and TPIC17-400K dataset. That means our approach provides a more accurate model for popularity prediction.

\subsection{Temporal Context Analysis}
Here we analyze different types of temporal context in terms of using only one of them in our prediction model on TPIC17-100K. There can be different variants of distinct temporal context type settings $TC_{t_{unit}}$-$t_r$, based on the type of time unit $t_{unit}$ and the temporal range of context $t_r$ chosen. Generally, short-term settings are applied for NTC types, which ranges from 0.5 day to 1 week in our experiment. Simultaneously, PTC types are applied to explore long-term patterns for popularity evolving, which ranges from 3 days to 1 month. The time unit $t_{unit}$ control the time-scale of NTC and PTC. For instance, $TC_{1P}$-$1D$ of NTC means to construct NTC from the data in previous 1 day range using the time-scale on period of day. Another case is $TC_{1D}$-$3W$ of PTC, which means the range of PTC for prediction target is in the same weekday of previous 3 weeks. 

 The results of using different types of NTC and PTC in our model are shown in Table \ref{comparison of contexts}, where the highest SRC is 0.5875 when using $TC_{1P}$-$1D$ of NTC. Overall, the results using NTC settings are better than PTC settings, which reveals that there are obvious short-term patterns in the 100K dataset of Flickr. Next, we analyze the improvements of performance in terms of NTC and PTC, respectively. Compared with the performance of DTCN on same data in Table \ref{SRC Comparison}, the performances of using a single temporal context (NTC or PTC) in Table \ref{comparison of contexts} drop down from 0.5990 to 0.5875 and 0.5745. This finding illustrates that prediction using both of NTC and PTC on same data performs more accurately than using one of them only, and our method can incorporate them together effectively.

{
\begin{table}[tb]
\centering
\caption{Prediction performances of context type NTC and PTC (metric: Spearman Ranking Correlation).}
\label{comparison of contexts}
\begin{tabular}{cccc}
\hline
\multicolumn{2}{c|}{\textbf{NTC}}      & \multicolumn{2}{c}{\textbf{PTC}} \\ \hline
\textbf{Context Type} & \multicolumn{1}{c|}{\textbf{SRC}}        & \textbf{Context Type}            & \textbf{SRC}        \\ \hline
$TC_{1P}$-$0.5D$    & \multicolumn{1}{c|}{0.5607}          & $TC_{1P}$-$3D$                 & 0.4975              \\
$TC_{1P}$-$1D$     & \multicolumn{1}{c|}{\textbf{0.5875}} & $TC_{1P}$-$5D$                 & \textbf{0.5745}              \\
$TC_{1D}$-$3D$      & \multicolumn{1}{c|}{0.5549}          & $TC_{1D}$-$3W$                 & 0.5365              \\
$TC_{1D}$-$7D$      & \multicolumn{1}{c|}{0.5784}          & $TC_{1D}$-$4W$                 & 0.5485              \\ \hline
\end{tabular}
\end{table}
}

\section{Conclusions and Future Work}
This paper proposes to learn consecutive temporal coherence of temporal context for predicting social media popularity over time, which comprises of three components. Firstly, we design a joint embedding network for multi-modal feature representation. Then, we propose and utilize two types of temporal context alignment to learn sequential popularity in short-term and long-term popularity fluctuations. Furthermore, we provide a temporal attention mechanism for predicting popularity at multiple time-scales. To evaluate our approach, we constructed a publicly available dataset with user-photo sequence data. Experimental results show that our prediction network achieves promising performances and outperforms the state-of-the-art deep prediction algorithms by 5.79\%--44.86\% relative improvements on TPIC17. The technique serves a deep prediction framework for sequential popularity prediction, and our paper along with the released dataset further helps promote the research. 

There are several possible directions for future investigations on social popularity prediction. One is considering the social network structure (and inherent social network analysis techniques) to improve the popularity prediction. Another open question is to exploit impact of most influential users in the social network.

\section*{Acknowledgements}
This work was supported in part by the National Key Research and Development Program of China under Grant 2016YFB0800403 and the National Nature Science Foundation of China (61525206, 61571424).



\balance

\bibliographystyle{named}
\bibliography{DeepPopularityPrediction}  

\end{document}